\documentclass[12pt]{iopart}
\usepackage{amssymb,graphicx}

\begin{document}

\title[Comment]{Comment on `Monte Carlo simulation study of the
two-stage percolation transition in enhanced binary trees'}
\author{Seung Ki Baek$^1$, Petter Minnhagen$^1$, Beom Jun Kim$^2$}
\address{$^1$ Department of Theoretical Physics, Ume{\aa} University, 901 87 Ume{\aa}, Sweden}
\address{$^2$ BK21 Physics Research Division and Department of Energy Science,
Sungkyunkwan University, Suwon 440-746, Korea}
\ead{beomjun@skku.edu}

\begin{abstract}
The enhanced binary tree (EBT) is a nontransitive graph which has
two percolation thresholds $p_{c1}$ and $p_{c2}$ with $p_{c1}<p_{c2}$.
Our Monte Carlo study implies that the second threshold $p_{c2}$ is
significantly lower than a recent claim by Nogawa and Hasegawa (J. Phys. A:
Math. Theor. {\bf 42} (2009) 145001). This means that $p_{c2}$ for the EBT does
not obey the duality relation for the thresholds of dual graphs
$p_{c2}+\overline{p}_{c1}=1$ which is a property of a transitive, nonamenable,
planar graph with one end. As in regular hyperbolic lattices, this relation
instead becomes an inequality $p_{c2}+\overline{p}_{c1}<1$.
We also find that the critical behavior is well described by the scaling
form previously found for regular hyperbolic lattices.
\end{abstract}

\maketitle

Recently, Nogawa and Hasegawa~\cite{noha} reported the two-stage percolation
transition on a nonamenable graph which they called the enhanced binary tree
(EBT). While the first transition had little ambiguity, they mentioned that
the behavior at the second threshold did not look like a usual continuous
phase transition.

\begin{figure}
\begin{center}
\includegraphics[width=0.45\textwidth]{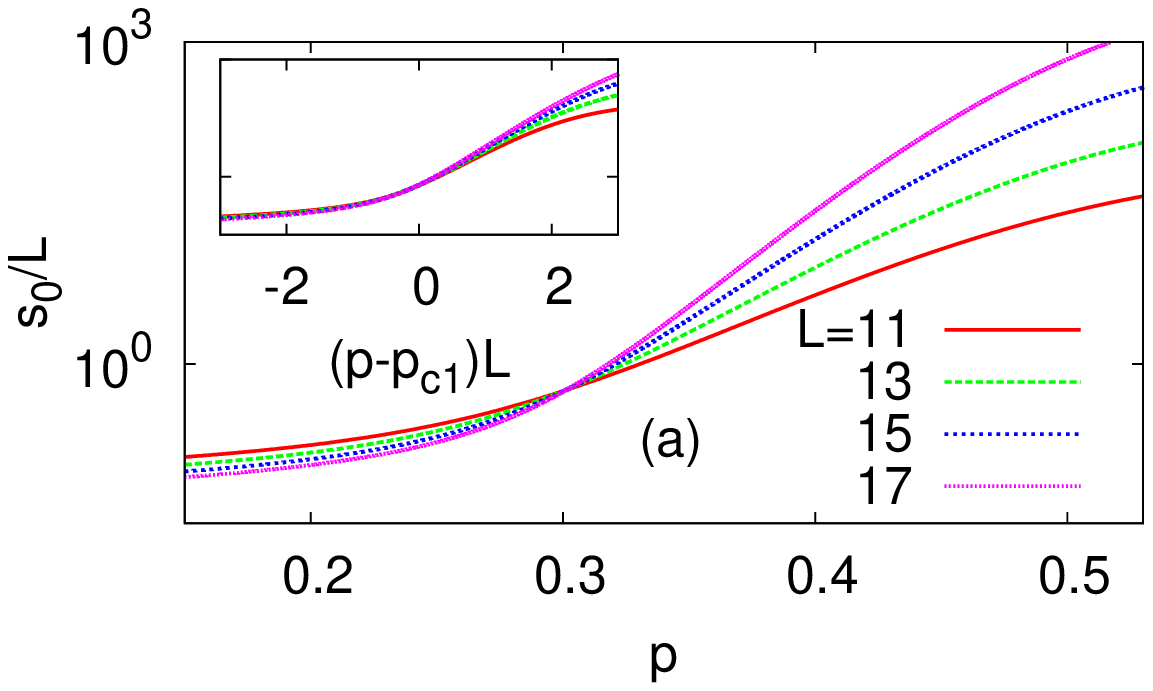}
\includegraphics[width=0.45\textwidth]{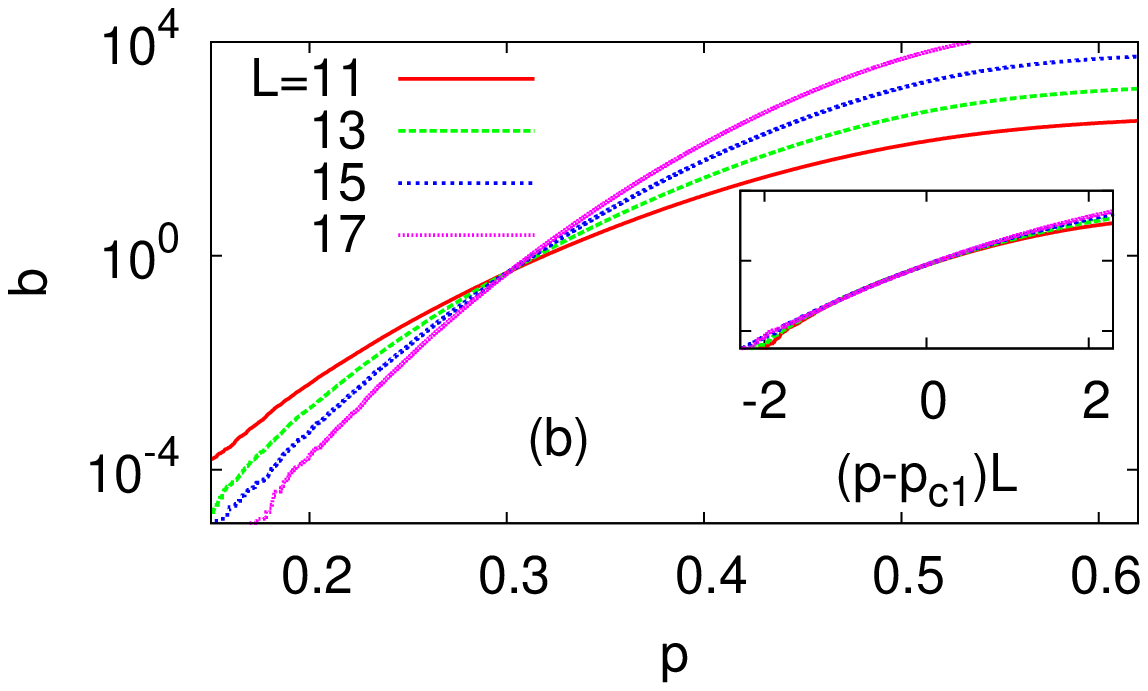}
\end{center}
\caption{({\it a}) Mass of the root cluster, $s_0$, divided by $L$, the number of
generations in the EBT. The crossing point indicates the first percolation
transition point, $p_{c1}$, where an unbounded cluster emerges. Inset:
Scaling collapse by equation~(\ref{eq:pc1}) with $p_{c1}=0.304$ found in
\cite{noha}. ({\it b}) The number of boundary points connected to the root
node, denoted as $b$, also shows a crossing point at $p=p_{c1}$.
Inset: Scaling collapse by equation~(\ref{eq:b}) with the same $p_{c1}$ as above.
}
\label{fig:pc1}
\end{figure}

A quantity of interest was the mass of the root cluster, denoted as $s_0$,
where the root cluster was defined as the one including the root node of
the EBT. Using this observable, we briefly check the first transition point,
$p_{c1}$, where an unbounded cluster begins to form. As in \cite{baek},
we have used the Newman-Ziff algorithm~\cite{nz1,nz2} and taken averages
over $10^6$ samples throughout this work. The number of generations, $L$, of
the EBT defines a typical length scale of the system, and \cite{noha}
showed the finite-size scaling of $s_0$ as
\begin{equation}
s_0/L \propto \tilde{f_1}[(p-p_{c1}) L^{1/\nu}],
\label{eq:pc1}
\end{equation}
with $\nu = 1$. Figure~\ref{fig:pc1}({\it a}) confirms both of the percolation
threshold $p_{c1}$ and the scaling form, equation~(\ref{eq:pc1}).
Equivalently, one can measure $b$, the number of boundary
points connected to the root node, which becomes finite above $p_{c1}$
as shown in figure~\ref{fig:pc1}({\it b}). It also scales as
\begin{equation}
b \propto \tilde{f_2}[(p-p_{c1}) L^{1/\nu}],
\label{eq:b}
\end{equation}
with the same exponent $\nu$.
Comparing this with \cite{baek}, we see that the
percolation transition in the EBT at $p=p_{c1}$ belongs to the same universality
class as that of regular hyperbolic lattices.
One may argue that this scaling form actually corresponds to the case of
Cayley trees~\cite{baek}. The convincing results in figure~\ref{fig:pc1}
imply that the estimation in \cite{noha} for the dual of the EBT,
$\overline{p}_{c1} = 0.436$, is also correct.

\begin{figure}
\begin{center}
\includegraphics[width=0.45\textwidth]{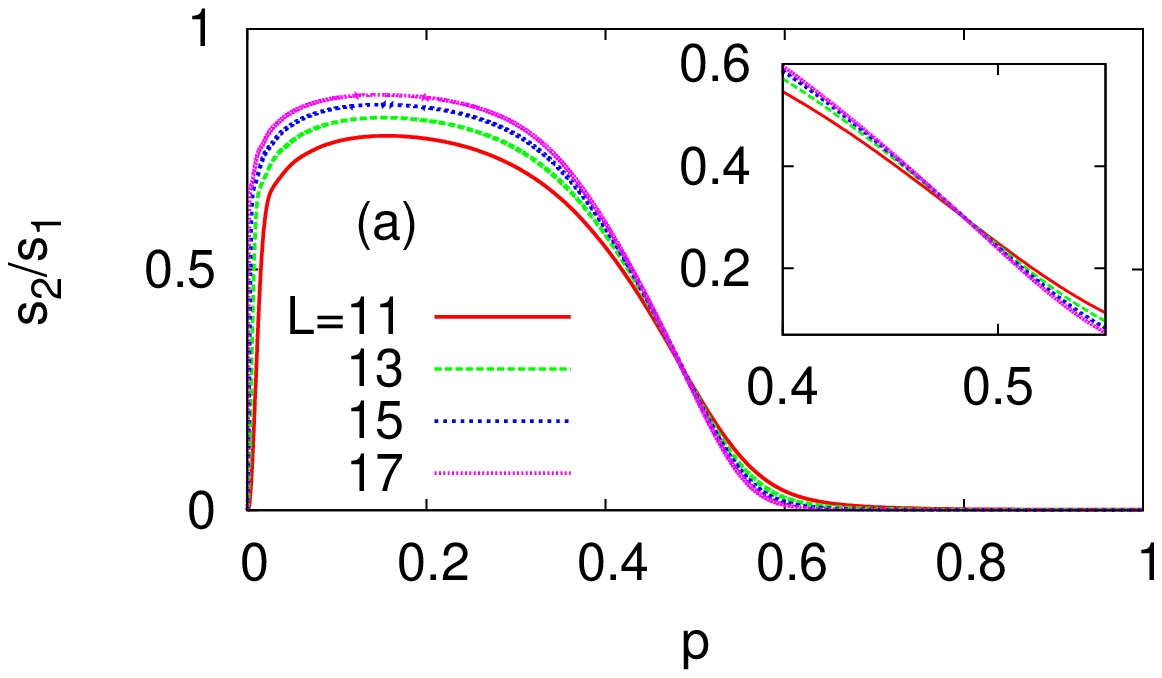}\\
\includegraphics[width=0.45\textwidth]{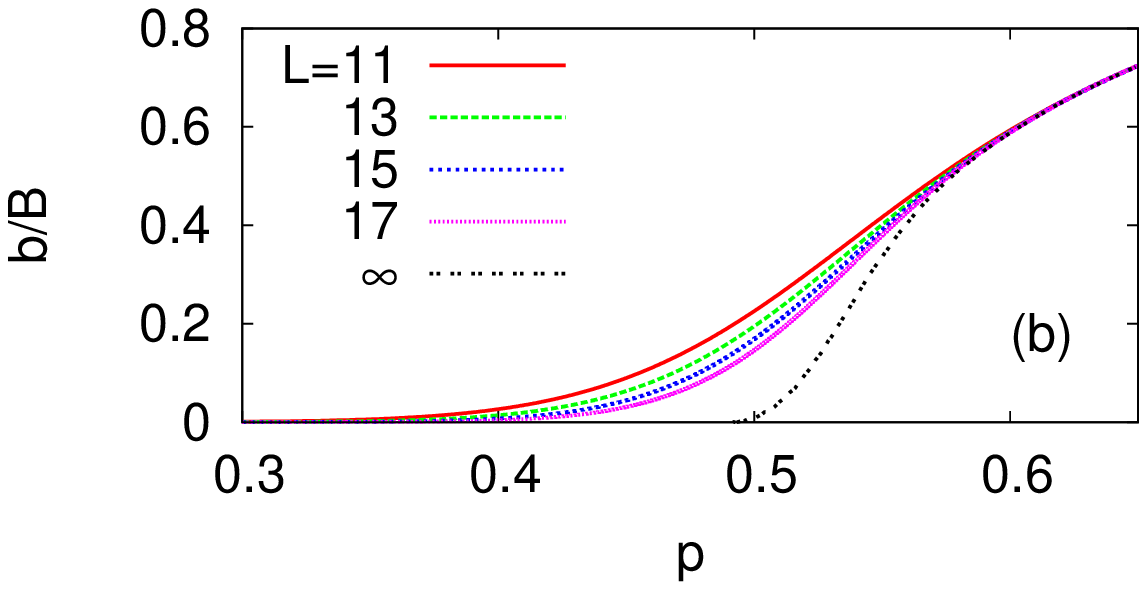}
\includegraphics[width=0.45\textwidth]{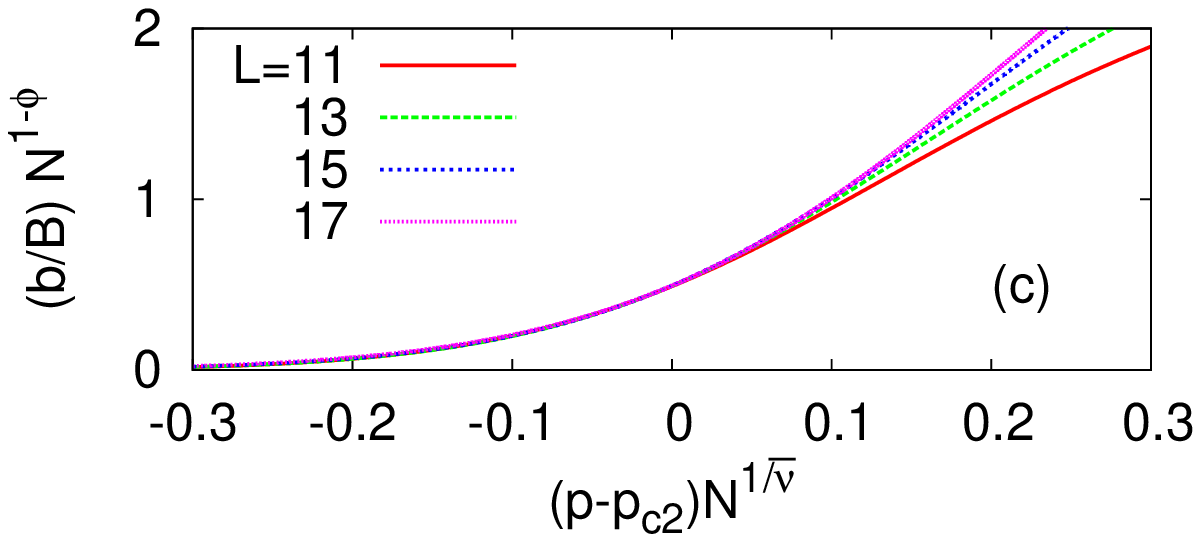}
\end{center}
\caption{({\it a}) Ratio between the second largest cluster mass, $s_2$, and
the first largest one, $s_1$. The crossing point lies at $p \approx 0.48$,
which implies that only one cluster will remain dominant in the
system.
({\it b}) Number of boundary points connected to the root node, $b$, divided by
the total number of boundary points, $B$. The dotted black curve marked by
$\infty$ indicates the extrapolation result from equation~(\ref{eq:ext}).
({\it c})
Scaling collapse according to equation~(\ref{eq:scale}), where we set
$p_{c2}=0.48$, $\phi=0.84$, and $1/\bar\nu = 0.12$.}
\label{fig:pc2}
\end{figure}

On the other hand, the second percolation transition at $p=p_{c2}$ indicates
uniqueness of the unbounded cluster. We have thus employed a direct
observable to detect this transition, i.e., the ratio between the first and
second largest cluster masses~\cite{baek}. The idea is that even the second
largest cluster would become negligible if there can exist only one unique
unbounded cluster. Measuring $s_2/s_1$ in the EBT, where $s_i$ means the $i$th
largest cluster mass, we have found the second transition at $p_{c2} \approx
0.48$ (figure~\ref{fig:pc2}({\it a})), certainly lower than Nogawa and Hasegawa's
estimation, $p=0.564$.

As an alternative quantity for $p_{c2}$, we divide $b$ by the number of all
the boundary points, $B$. This fraction $b/B$ is supposed to become finite
above $p_{c2}$~\cite{baek}. Based on the Cayley tree result~\cite{baek}, we
have assumed that as the system size $N$ varies, one can write down the
following asymptotic form:
\begin{equation}
b/B \sim c_1 N^{\phi-1} + c_2,
\label{eq:ext}
\end{equation}
with some constants $c_1$ and $c_2$ and an exponent $\phi$. From the
finite-size data, we extrapolate the large-system limit by
equation~(\ref{eq:ext}), which suggests $p_{c2} \approx 0.49$
(figure~\ref{fig:pc2}({\it b})). This is very close to the estimation above from
$s_2/s_1$.
Moreover, in accordance with equation~(\ref{eq:ext}), we have suggested the
following scaling hypothesis to describe the critical behavior at this
transition point~\cite{baek}:
\begin{equation}
b/B \propto N^{\phi-1} \tilde{f_3}[(p-p_{c2}) N^{1/{\bar\nu}}],
\label{eq:scale}
\end{equation}
with an exponent $\bar\nu$. Applying this hypothesis to EBT data, we see
that $\phi=0.84$ and $1/\bar\nu=0.12$ give a good fit
(figure~\ref{fig:pc2}({\it c}))
with the same value of $p_{c2} = 0.48$,
where the numeric values of the scaling exponents are again consistent
with \cite{baek}.

\begin{figure}
\begin{center}
\includegraphics[width=0.45\textwidth]{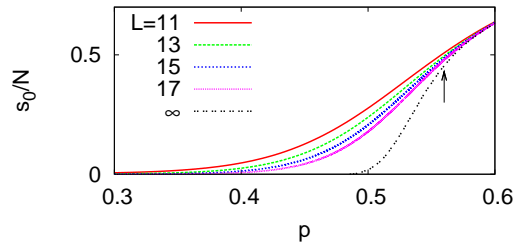}
\end{center}
\caption{Mass fraction of the root cluster.
The extrapolation result is represented by the dotted black curve named as
$\infty$.
The arrow indicates $p=0.564$, predicted as the transition point
in \cite{noha}.}
\label{fig:s0}
\end{figure}

To make a direct comparison to the observation in \cite{noha},
we have also calculated the mass fraction of the root cluster, $s_0/N$, as a
function of $p$. As above, performing extrapolation to the large-system
limit, we see that this quantity becomes positive finite at $p \gtrsim 0.49$
(figure~\ref{fig:s0}).

All of these observations suggest that the predicted value of $p_{c2}$ in
\cite{noha} is too high, and it seems that this overestimation
led them to consider `discontinuity' since $s_0/N$ became already
so large at that point as shown in figure~\ref{fig:s0}.

\begin{figure}
\begin{center}
\includegraphics[width=0.6\textwidth]{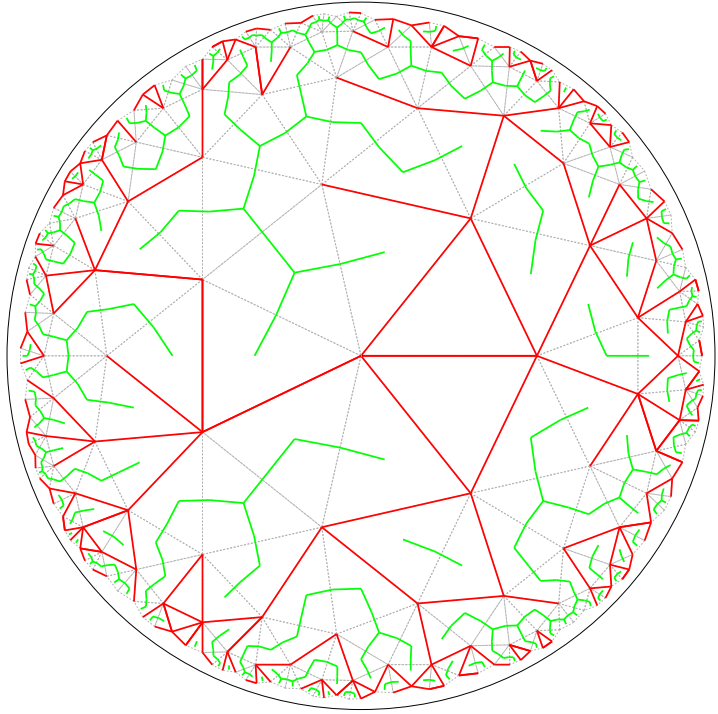}
\end{center}
\caption{Visualization of a triangular hyperbolic lattice projected on the
Poincar\'e disk, where the maximum length from the origin is chosen to be
$L=4$. Bonds are randomly occupied with probability of $p=0.42$, which
are colored red, while only the rest of them appear as occupied in the
dual lattice, as colored green, so that the dual probability corresponds to
$\overline{p} = 1-p = 0.58$.
Note that $p$ lies between $p_{c2}$ and
$1-\overline{p}_{c1}$, since this structure has $p_{c2} \approx 0.37$ and its
dual has $\overline{p}_{c1} \approx 0.53$, according to \cite{baek}.
While most clusters have been already absorbed into the largest red cluster,
many of green clusters still have radii comparable to $L$ since
$\overline{p}_{c1} < \overline{p} < \overline{p}_{c2} \approx 0.72$.}
\label{fig:vis}
\end{figure}

Finally, even though our estimation suggests such a different $p_{c2}$
that $p_{c2} + \overline{p}_{c1} < 1$, we note that it does not violate the
duality relation proved in \cite{ben} for a transitive, nonamenable,
planar graph with one end: As Nogawa and Hasegawa correctly pointed
out~\cite{noha}, the EBT does not possess transitivity.
The inequality $p_{c2} + \overline{p}_{c1} < 1$ was explicitly verified for a
pair of hyperbolic dual lattices $\{7,3\}$ and $\{3,7\}$ in
\cite{baek}. This inequality means the existence of a narrow region of
$p$ between $p_{c2}$ and $1-\overline{p}_{c1}$, where one would find a unique
unbounded cluster in a given graph whereas infinitely many unbounded
clusters in its dual graph. Such a region does not exist for a transitive
case~\cite{ben}. A typical state in this region is illustrated in
figure~\ref{fig:vis}, which shows a situation with many unbounded clusters of
radii comparable to $L$ at the same time as a single unbounded cluster
occupies the dominant part of the dual graph.

\ack
SKB and PM acknowledge the support from the Swedish Research Council
with the Grant No. 621-2002-4135. BJK was supported by
the Korea Research Foundation Grant funded by the Korean
Government (MOEHRD) with Grant No. KRF-2007-313-C00282.
This research was conducted using the resources of High Performance
Computing Center North (HPC2N).


\section*{References}

\begin{thebibliography}{1}

\bibitem{noha}
Nogawa T and Hasegawa T 2009
{\it J. Phys. A: Math. Theor.} {\bf 42} 145001

\bibitem{baek}
Baek S K, Minnhagen P and Kim B J 2009
{\it Phys. Rev. E} {\bf 79} 011124

\bibitem{nz1}
Newman M E J and Ziff R M 2000
{\it Phys. Rev. Lett.} {\bf 85} 4104

\bibitem{nz2}
Newman M E J and Ziff R M 2001
{\it Phys. Rev. E} {\bf 64} 016706

\bibitem{ben}
Benjamini I and Schramm O 2000
{\it J. Am. Math. Soc} {\bf 14} 487

\end{thebibliography}
\end{document}